This work has been accepted for publication at SmartNets 2026

# StormWave: An Open-Source Portable SDR Platform for Over-the-Air Resilience Evaluation of Terrestrial and Aerial Communications

Yuqing Cui[1], Zhaoxi Zhang[1], Sidharth Santhi Nivas[1], Prem Sagar Pattanshetty Vasanth Kumar[2], Maxwell McManus[2], Chenzhi Zhao[1], Guanying Sun[2], Nicholas Mastronarde[2], George Sklivanitis[3], Dimitris A. Pados[3], Elizabeth Serena Bentley[4], Zhangyu Guan[1]
[1]Dept. CSE, University of Minnesota-Twin Cities, USA; [2]Dept. EE, University at Buffalo, USA; [3]Dept. EECS, Florida Atlantic University, USA; [4]U.S. Air Force Research Laboratory (AFRL)
Email: {zha00410, santh055, zhao2305, zguan}@umn.edu, elizabeth.bentley.3@us.af.mil, {premsaga, memcmanu, gsun4, nmastron}@buffalo.edu, {gsklivanitis, dpados}@fau.edu

***Abstract*—This paper presents *StormWave*, an open-source, portable software-defined Radio Frequency (RF) interference generation and monitoring platform designed for realistic field-based evaluation of the resilience of wireless communication systems. StormWave enables seamless composition and runtime switching among a wide range of narrowband and wideband waveforms, while supporting multiple digital modulations, adaptive coding, and multi-radio orchestration with real-time spectrum visualization. We evaluate the effectiveness of StormWave through both outdoor ground and air-to-air (A2A) experiments. Ground experiments demonstrate clear waveform- and modulation-dependent interference effects under realistic propagation conditions, while A2A experiments reveal pronounced distance-dependent constellation distortion and access-symbol degradation under active interference. The StormWave source code will be released to the community, with the expectation that StormWave will be used as a flexible, extensible, and field-ready platform for systematically validating interference resilience of wireless systems under realistic operating conditions.**

## I. Introduction

The ability to generate and inject reproducible RF interference has become a critical capability for validating the resilience of modern wireless communication systems [1]–[3]. By introducing structured and repeatable interference conditions, researchers can systematically analyze system vulnerabilities, quantify performance degradation, and design adaptive protocols that operate reliably in hostile or congested spectrum environments. This capability is particularly important for wireless systems in which robustness under interference is a primary concern, such as navigation systems [4], [5] and unmanned aerial system (UAS) [6], [7].

Early interference-generation platforms were typically built around fixed or minimally configurable signal generators and multi-component hardware architectures, which limited their suitability for agile experimentation or field deployment [8]. Subsequent efforts have explored a range of interference strategies, including broadband and barrage-style interference to study denial-of-service (DoS) resilience. Wideband noise has been shown to severely degrade GPS reception [9], with similar effects demonstrated for aerial navigation systems using SDR-based emitters [6]. Programmable SDR platforms have also enabled structured broadband interference against Wi-Fi systems, achieving effective disruption at the cost of poor spectral efficiency [10]. While effective, these approaches typically emphasize disruption strength rather than waveform agility, experimental controllability, or portability.

In parallel, narrowband, tone-based, and sweeping interference techniques have proven effective against narrowband receivers and aerial control links. For example, protocol-aware tone and sweep methods degrade aerial links based on Frequency Hopping Spread Spectrum (FHSS) and Direct Sequence Spread Spectrum (DSSS), with reduced collateral impact [7], while frequency-sweeping and comb-based approaches can induce acquisition failures in Global Navigation Satellite System (GNSS) receivers [4], [5]. Beyond continuous interference, sparse spectral and temporal strategies have been explored to improve efficiency. Comb- and pulse-based vulnerabilities in Orthogonal Frequency-Division Multiplexing (OFDM) systems have been analyzed theoretically [11], while learning-based detection and evaluation of pulse, barrage, and protocol-aware interference have been studied in Unmanned Aerial Vehicle (UAV) and OFDM links [12]. More recent work examines vulnerabilities in joint sensing-and-communication OFDM systems [13]. Adaptive and protocol-aware interference techniques further leverage waveform knowledge to improve effectiveness, including methods based on multi-agent reinforcement learning [14] and deep learning [15].

However, despite our prior prototype demonstrating SDR-based RF generation [16], significant gaps remain between algorithmic or simulation-based studies and realistic, field-based evaluation. Many existing SDR-based interference platforms

This work was supported in part by the National Science Foundation (NSF) under Grant SWIFT-2229563 and CNS-2450418, and the U.S. Air Force Research Laboratory under Contracts FA8750-21-F-1012, FA8750-20-C1021 and FA8750-25-1-1000.

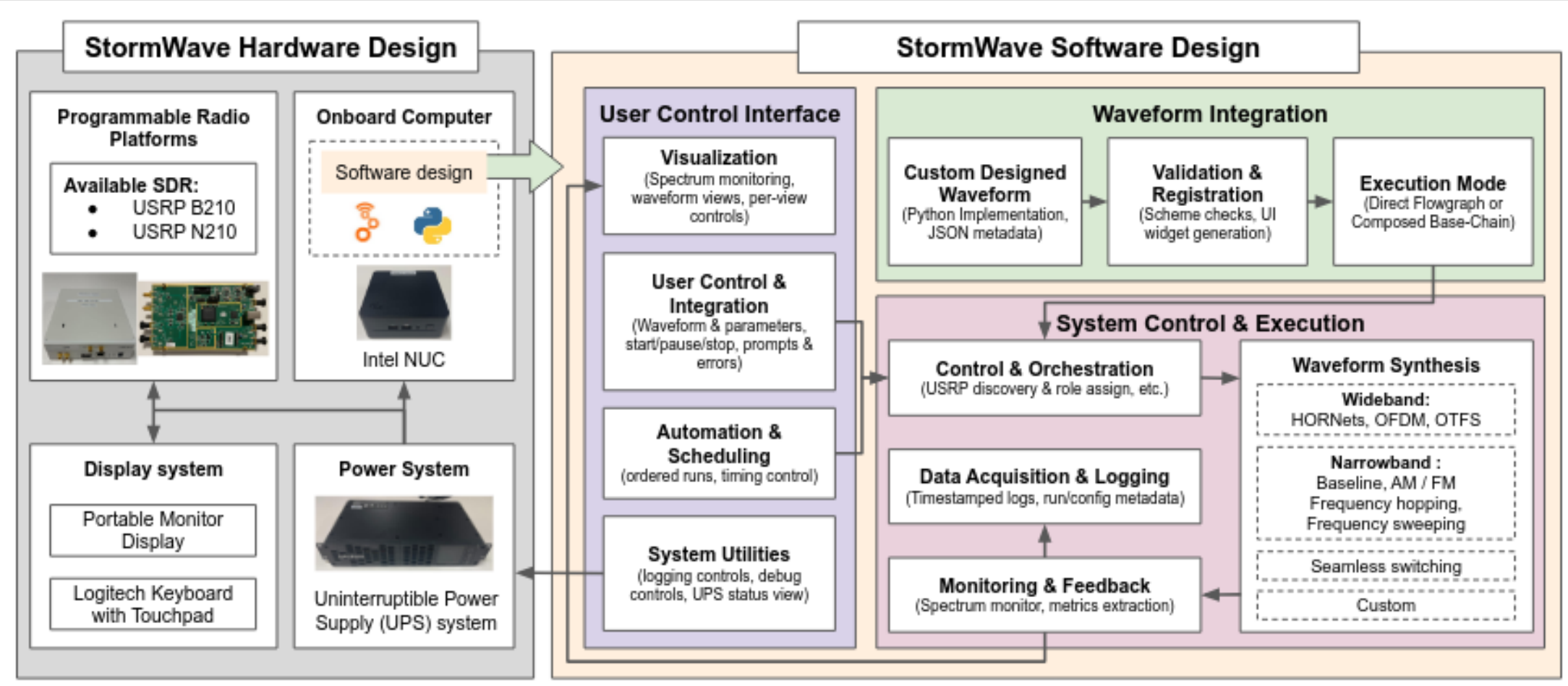

Fig. 1: Overall System Architecture of StormWave.

lack seamless runtime waveform switching, unified support for user-defined waveform integration, tightly coupled visualization and control, and automatic management of multiple transceivers. Practical considerations such as portability, environmental resilience, and continuous operation in disconnected field environments are rarely addressed.

**Contributions.** To bridge these gaps, we introduce *StormWave*, an open-source, portable, weatherproof, and low-cost software-defined RF interference generation and monitoring platform designed for comprehensive, field-based evaluation of wireless communication systems. StormWave enables users to flexibly test a wide range of built-in and custom-developed waveforms, supports seamless runtime switching between heterogeneous interference profiles, provides intuitive real-time user control with integrated spectrum visualization, and automatically manages available USRP devices for optimal operation. With its modular architecture, long-duration power system, and mobility-ready design, StormWave is explicitly designed to support repeatable experiments in indoor, as well as dynamic outdoor and airborne environments.

The StormWave source code will be made available to the community upon publication of the article. By releasing StormWave as an open extensible platform, our aim is to enable reproducible evaluation of interference resilience of wireless systems under realistic operating conditions.

## II. StormWave System Design

The primary objective of StormWave is to enable simple, rapid and repeatable evaluation of wireless systems in diverse interference conditions, ranging from basic narrowband emitters to advanced wideband and protocol-aware waveforms. To this end, the StormWave design prioritizes i) modular waveform composition and extensibility through user-uploaded implementations; ii) seamless runtime reconfiguration during active operation without hardware re-initialization; iii) integrated sensing, visualization, and logging for experiment traceability; and iv) robust operation in disconnected or power-constrained environments.

A secondary objective of StormWave is to reduce the operator burden and mitigate configuration errors during complex experimental campaigns. StormWave adopts a Graphical User Interface (GUI)-centric control model that supports automatic device discovery, parameter validation, synchronized visualization, and integrated power management. This design allows users to focus on experimental execution rather than low-level system configuration. As illustrated in Fig. 1, the StormWave architecture tightly integrates hardware and software components into a unified, field-deployable platform.

### A. Hardware Design

StormWave consists of four primary hardware modules: an onboard computing unit, programmable radio platforms, a power system, and an integrated display system.

The onboard computing unit acts as the central control and orchestration module of StormWave. It is implemented using an Intel NUC with a multi-core Intel i7-class CPU and 32 GB of RAM, providing sufficient capacity to support real-time waveform execution, spectrum monitoring, seamless waveform switching, visualization, and metadata logging. To enable field deployment and unattended operation, the computing unit supports remote access via standard IP-based interfaces, including secure shell and NX-based remote desktop protocols over Ethernet or Wi-Fi. As shown in Fig. 1, the onboard computer functions as the hub connecting the programmable radio platforms, display system, and power system.

The programmable radio platforms consist of two USRP-class SDRs: a USRP B210 and a USRP N210 with a CBX daughterboard. Together, these radios provide continuous frequency coverage from approximately 70 MHz to 6 GHz, supporting a wide range of narrowband and wideband interference waveforms. USRP devices are selected for their mature UHD (i.e., USRP Hardware Driver) driver support, deterministic buffering behavior, and stable timing control, which are critical for multi-radio operation and seamless waveform switching. While other low-cost SDR platforms (e.g., HackRF One [6], [10], BladeRF [7], LimeSDR [17], and PlutoSDR [5]) are commonly used in interference studies, their more limited dynamic range, driver maturity, or interface stability make them less suitable for StormWave's concurrent transmission and monitoring requirements.

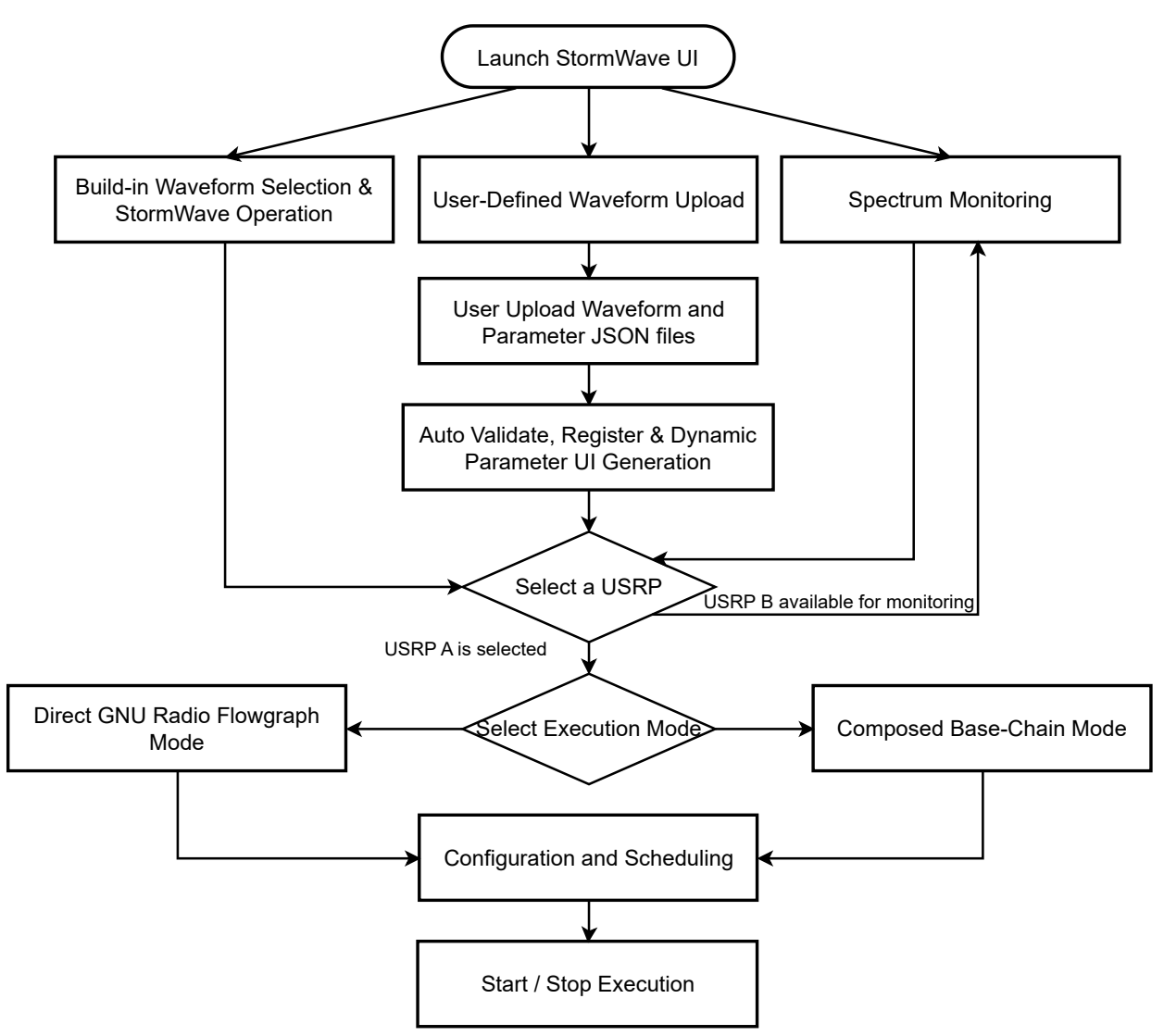

Fig. 2: StormWave user interaction design flowchart.

As illustrated in Fig. 2, one of the two SDR radios, e.g., the USRP N210 is assigned to interference transmission and interfaces with the onboard computing unit over Gigabit Ethernet, providing a stable, high-throughput data path for continuous waveform generation. The USRP B210 is dedicated to continuous spectrum monitoring and connects via USB 3.0, enabling portable deployment and low-latency capture of RF samples. Both radios are configured and controlled through UHD, allowing coordinated operation and dynamic role assignment without restarting the hardware.

The power system is built around a CyberPower PFC-sinewave uninterruptible power supply (1500 VA / 1000 W), providing up to two hours of tested runtime under field conditions. This enables continuous operation during outdoor experiments without reliance on external power sources. Finally, the display system includes an integrated portable monitor and a compact keyboard with touchpad, allowing local configuration, monitoring, and control even in the absence of network connectivity.

## B. Software Design

As illustrated in Fig. 1, StormWave adopts a three-layer modular software architecture consisting of the User Control Interface, Waveform Integration, and System Control & Execution layers, as described as follows.

*1) User Control Interface Layer:* This layer is implemented as a PyQt-based graphical user interface that provides centralized control over waveform selection, parameter configuration, spectrum visualization, power monitoring, and data collection automation. Upon startup, the GUI automatically discovers all connected USRPs and displays them for user assignment as transmission or spectrum-monitoring devices. As shown in Fig. 1, the control interface is further organized into four functional blocks: User Control & Integration, Visualization, System Utilities, and Automation & Scheduling.

The *User Control & Integration Block* manages waveform selection, execution control (start, pause, and stop), and user interactions, while performing parameter validation using metadata and JSON configuration files. These metadata files are used to dynamically generate configuration widgets (e.g., frequency, gain, and modulation), ensuring consistent and error-resilient experiment setup. The *Visualization Block* provides real-time spectrum monitoring synchronized with waveform-domain views, enabling users to observe interference effects during transmissions. The *System Utilities Block* exposes logging controls, debugging utilities, and real-time UPS status, including battery capacity and estimated remaining runtime. Finally, the *Automation & Scheduling Block* enables orchestrated experiment execution with precise timing control, supporting repeatable runs with minimal operator intervention.

*2) Waveform Integration Layer:* The Waveform Integration Layer enables extensible interference waveform development through a well-defined API, bridging user interaction and runtime execution. New waveforms are added by supplying a Python implementation along with a metadata/JSON parameter description that declaratively specifies configurable waveform properties. Typical JSON fields include a waveform identifier (e.g., `waveform_name`), a waveform category (e.g., `narrowband` or `wideband`), and a set of parameter definitions such as `center_frequency`, `gain`, `modulation`, and `symbol_rate`, each annotated with type information, valid ranges, and units. These parameters are selected because they directly map to USRP configuration (frequency and gain), waveform structure (modulation and symbol rate), and base signal-processing requirements, allowing StormWave to configure heterogeneous waveforms in a uniform manner.

As shown in Fig. 1, new waveforms to be integrated are automatically processed by the *Validation & Registration Block*, which performs scheme checks, validates parameter definitions (e.g., integer, floating-point, or enumerated values), and range enforcement to prevent invalid RF configurations. Compatibility checks are also applied to ensure that the declared waveform category matches the supported base processing chain. Once registered, user-defined waveforms are integrated into the same control, visualization, and logging pipeline as built-in waveforms.

This layer supports two execution models selectable at runtime, as shown in Fig. 2, to accommodate waveforms with different complexity and integration requirements. *Mode 1 - Direct GNU Radio Flowgraph Mode:* In this mode, complex waveforms are implemented as self-contained GNU Radio flowgraphs and instantiated directly by the backend. This mode is intended for waveforms that require tightly coupled signal-processing pipelines, custom scheduling behavior, or nonstandard block compositions, and allows developers to retain full control over the GNU Radio execution graph with minimal abstraction overhead. *Mode 2 - Composed Base-Chain Mode:*

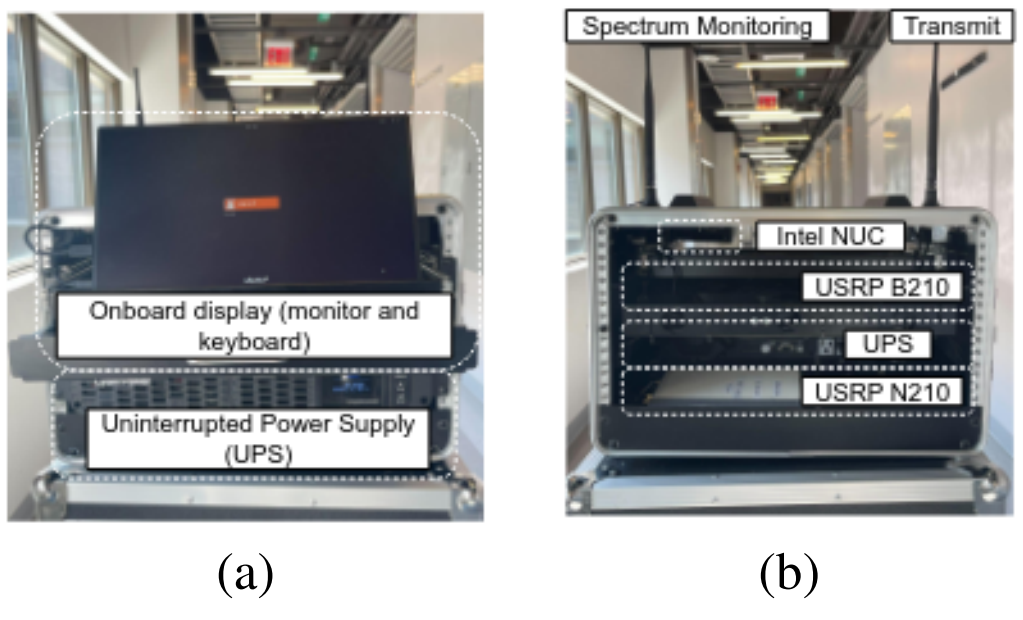

(a) (b)

Fig. 3: Snapshots of StormWave Prototype.

In this mode, waveforms are defined using lightweight classes that specify only waveform-specific signal generation and modulation logic; the backend then automatically composes these definitions with shared narrowband or wideband base signal-processing chains, including modulation, scaling, framing, and USRP binding. This mode aims to reduce duplication across waveform implementations and ensure consistent handling of hardware configuration and runtime control.

*3) System Backend Layer:* This layer is responsible for low-level runtime operation, coordinating hardware control, waveform execution, monitoring, and data logging. As shown in Fig. 1, this layer comprises four core blocks: *Control & Orchestration*, *Waveform Synthesis*, *Monitoring & Feedback*, and *Data Acquisition & Logging*. The *Control & Orchestration Block* manages USRP device discovery, role assignment, and configuration, and coordinates flowgraph lifecycle operations including start, pause, resume, and stop. This centralized control architecture enables seamless switching among heterogeneous waveforms during active transmission without requiring SDR hardware restarts. The *Waveform Synthesis Block* generates interference signals using both built-in and user-defined waveforms, supporting narrowband implementations (Baseline, AM/FM, frequency hopping, and frequency sweeping) as well as wideband designs (HORNet [18], OFDM, and OTFS [19]). The *Monitoring & Feedback Block* continuously captures spectrum measurements and extracts relevant runtime metrics, feeding this information back to the User Control Interface for synchronized real-time visualization and operator awareness. In parallel, the *Data Acquisition & Logging Block* records timestamped experiment data, configuration metadata, and execution logs, ensuring traceability and enabling repeatable post-processing and analysis.

## III. Prototyping and Experimental Evaluation

We have prototyped StormWave following the hardware and software architecture described in Section II. All hardware components are integrated into a rugged, weatherproof enclosure sized for easy transport and rapid deployment. The current StormWave prototype has overall dimensions of approximately 21.9 in (width) × 13.6 in (height) × 24.5 in (depth) and a total system weight of about 50 lb, allowing the platform to be transported and deployed by a single operator for field experimentation. In the current prototype, we implemented a set of interference waveforms spanning narrowband and wideband designs. Among these waveforms, the baseline waveform is a narrowband continuous transmission without time–frequency spreading and serves as a reference for minimal spectral overlap; HORNet is a wideband spread-spectrum waveform [18], OFDM employs multi-carrier modulation with orthogonal subcarriers, and OTFS spreads symbols across the joint time–frequency domain [19]. Fig. 3 shows snapshots of the resulting StormWave prototype. Next, we experimentally evaluate the effectiveness and flexibility of StormWave through a series of ground-based and air-to-air (A2A) experiments.

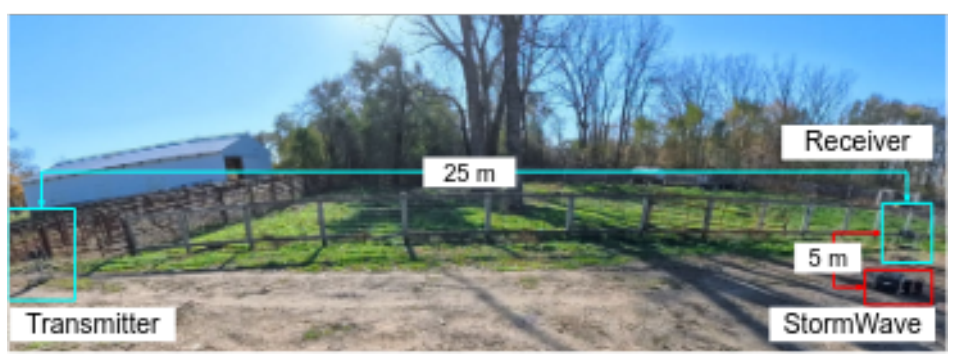

Fig. 4: Scenario 1: Ground field with surrounding blockages.

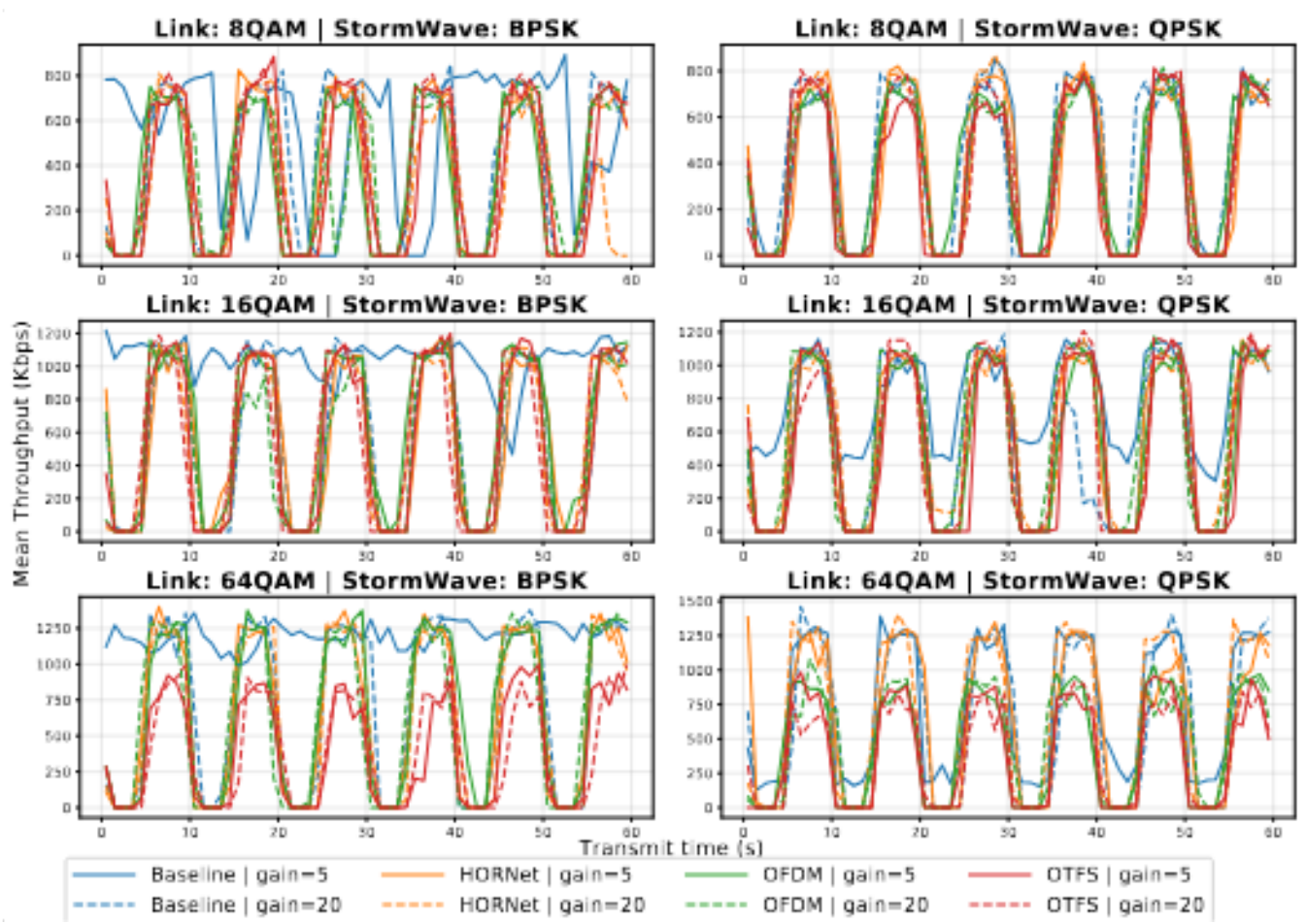

Fig. 5: Measured link throughput under StormWave-generated RF interference in Scenario 1.

**Ground Testing.** In the first experiment, we conduct ground-based outdoor field tests to capture the impact of key wireless channel effects, including path loss, obstruction, and multipath, on link behavior in the presence of StormWave-generated interference. For each scenario, the transmitter operates continuously for a total duration of one minute, while StormWave alternates between active and inactive states with a 5 s on/off duty cycle for each waveform, modulation and gain combination. StormWave employs BPSK and QPSK modulations across interference gain levels of 5–25 dB; Fig. 5 and Fig. 7 show representative results at 5 and 20 dB.

The test scenario is shown in Fig. 4, where the transmitter and receiver are separated by 25 m, with StormWave located 5 m from the receiver in a cluttered, multipath-rich environment. The measurement results are reported in Fig. 5. It can be seen that the baseline waveform causes only modest throughput degradation to the link, with throughput remaining above approximately 60–80% of nominal levels for both gain settings. In contrast, HORNet and OFDM induce periodic throughput collapses due to the 5 s on/off duty cycle, with QPSK generally causing deeper degradation than BPSK. For example, under QPSK interference at gain 20 dB, OFDM and OTFS throughput frequently drops below 20–30 Kbps. This behavior is amplified by rich multipath of the cluttered environment, which increases symbol overlap and interference coupling for higher-order modulations. Notably, even when the transmitter–receiver link uses HORNet, OTFS interference at gain = 5 dB can fully disrupt the link, demonstrating OTFS's strong time–frequency spreading and its effectiveness under close-range coupling. This experiment verifies StormWave can generate both narrowband and wideband RF interference waveforms, and that close-range coupling in multipath-rich environments significantly amplifies waveform-dependent interference effects, with wideband and time–frequency–spread waveforms causing the most severe degradation.

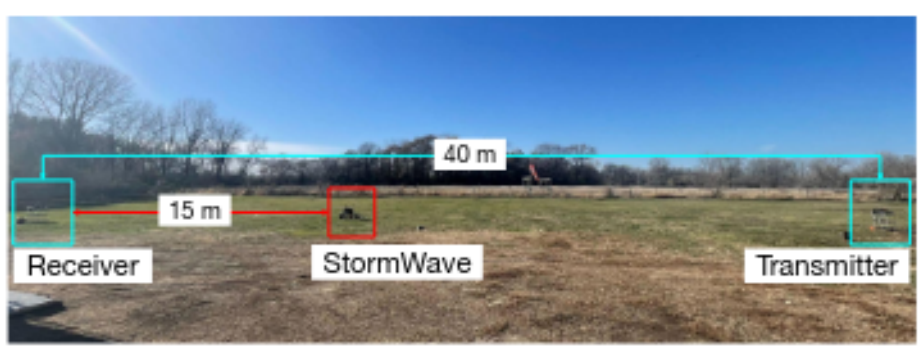

Fig. 6: Scenario 2: StormWave test in open field environment.

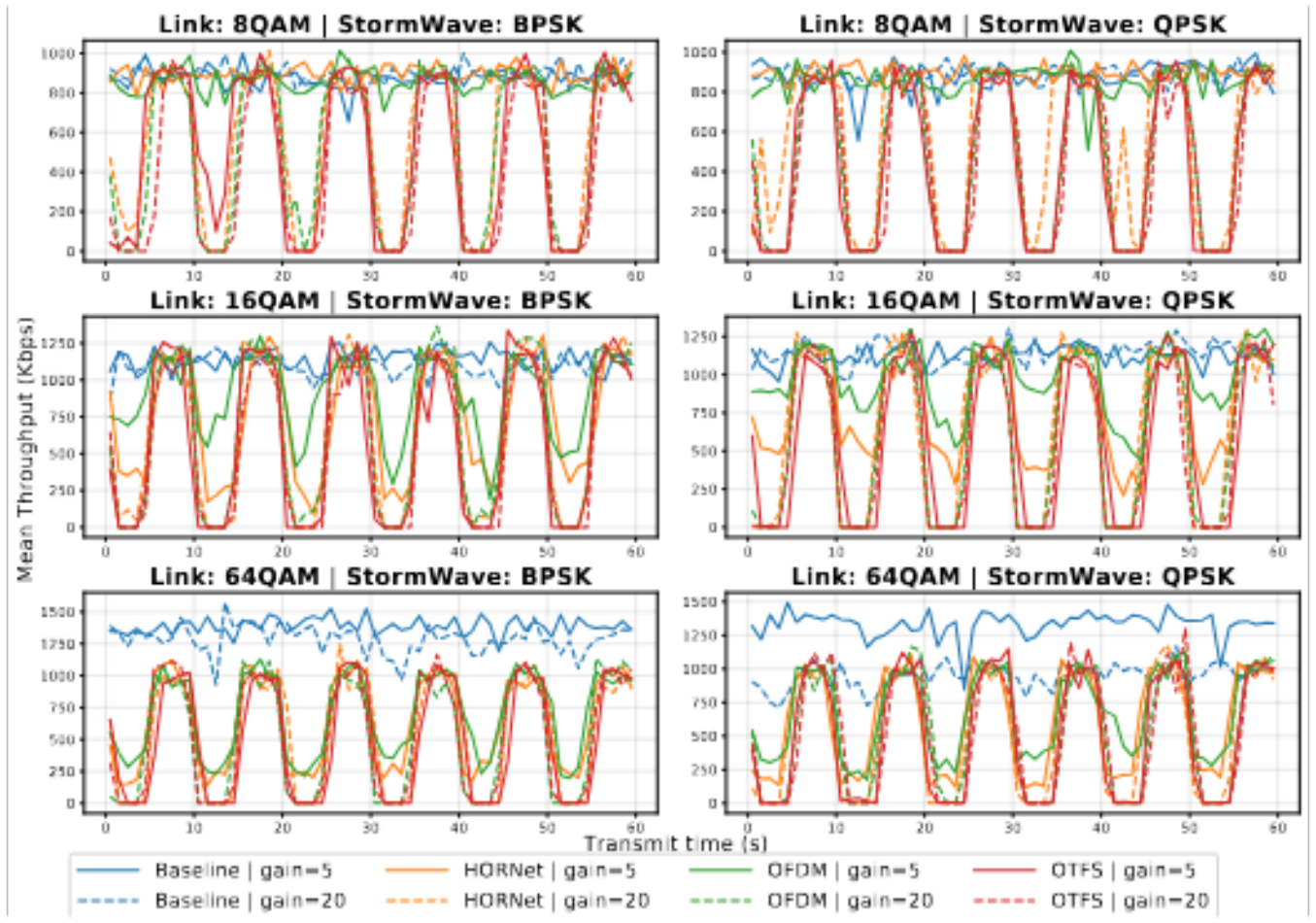

Fig. 7: Measured link throughput behavior in the presence of StormWave-generated RF interference in Scenario 2.

In the second experiment, as shown in Fig. 6, we consider a new scenario (Scenario 2) with a cleaner propagation environment. In this scenario, the transmitter–receiver separation increases to 40 m, and StormWave is repositioned to 15 m from the receiver. Compared to Fig. 5, throughput trends in Fig. 7 are stable and repeatable across cycles, reflecting reduced environmental complexity. Under these conditions, BPSK-based interference becomes more effective than QPSK, producing deeper (60–70%) and consistent throughput suppression across different waveforms (HORNet, OFDM, and OTFS). For instance, BPSK interference at gain 20 dB reduces average throughput to below 40% across multiple waveforms, whereas QPSK interference yields higher residual throughput. This inversion is consistent across receiver modulation schemes (8QAM, 16QAM, and 64QAM) and can be attributed to reduced multipath effects, where lower-order modulations, with higher energy concentration per symbol, dominate interference impact. HORNet remains the most resilient waveform across both scenarios, whereas OTFS continues to exhibit abrupt throughput collapses during active interference. Overall, these results indicate that increased separation and cleaner propagation conditions reduce interference variability and shift dominance toward lower-order modulation interference. A direct comparison of Fig. 5 and Fig. 7 further confirms that close proximity combined with multipath-rich environments significantly amplifies interference effectiveness, while increased separation primarily reduces variability and overall disruption.

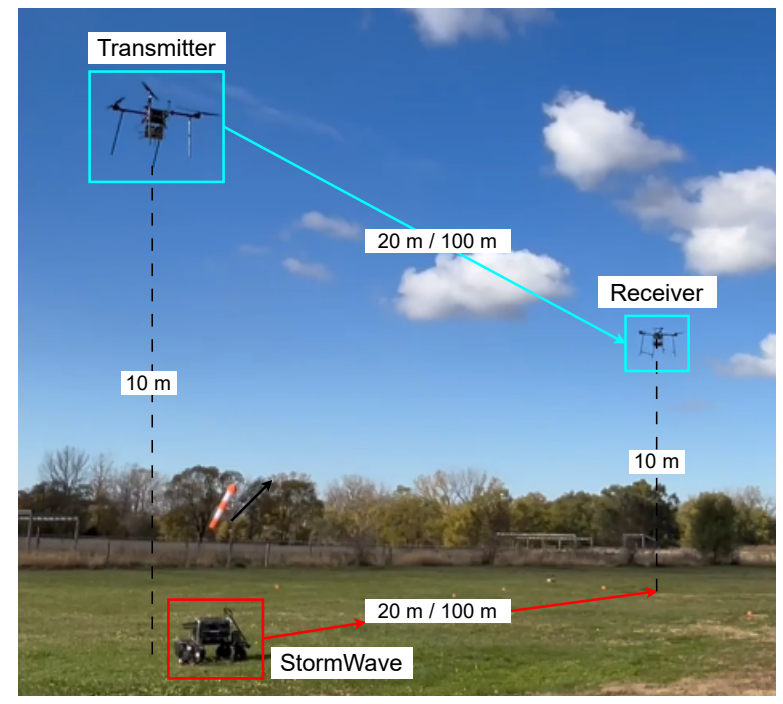

Fig. 8: Scenario 3: Air-to-Air Communication.

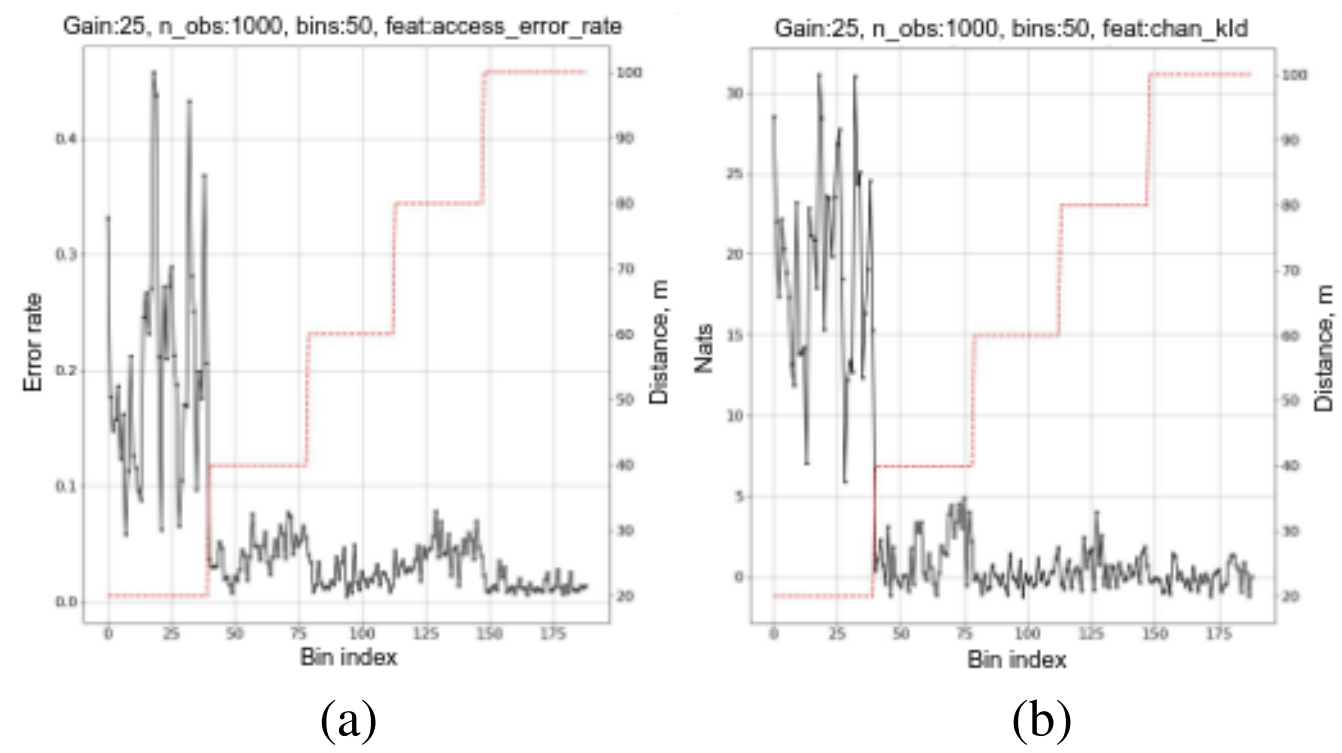

(a) (b)

Fig. 9: Air-to-Air I/Q results: (a) ASER and (b) KLD (nats); orange curve denotes distance (right y-axis).

**Air-to-Air Experiments.** To fully validate the effectiveness of StormWave against dynamically evolving aerial links, we conduct A2A experiments with both transmitter and receiver airborne and StormWave deployed on the ground (Fig. 8). The experiments are performed at an open outdoor test site, with the UAVs following controlled flight trajectories while the StormWave–receiver separation is varied. This setup captures mobility-induced and distance-dependent interference dynamics that are intrinsic to airborne communication scenarios.

Additionally, we evaluate StormWave's impact on the A2A link using raw baseband I/Q snapshots collected at the receiver UAV. This enables access-symbol error rate (ASER) and KL divergence (KLD) analysis from received I/Q samples under interference that remains meaningful even when higher-layer decoding fails. ASER is computed via symbol decision errors, and KLD measures distribution divergence between interfered and reference I/Q samples. Fig. 9 reports ASER (Fig. 9(a)) and KLD (Fig. 9(b)). At short separations of 20–40 m, the link is highly unstable under interference, with ASER frequently exceeding 20–40% and KLD values above 25 nats, indicating measurable degradation relative to the interference-free reference. As separation increases to 60–100 m, both metrics decrease and converge: ASER falls below 5%, and KLD approaches zero with reduced variance, reflecting stable and predictable link behavior. Notably, link stability improves with distance, indicating that close-range interference dominance outweighs path-loss effects in this regime. These results demonstrate a clear transition from interference-dominated operation to near-baseline behavior and validate the use of I/Q-based metrics for airborne evaluation when throughput measurements are unavailable.

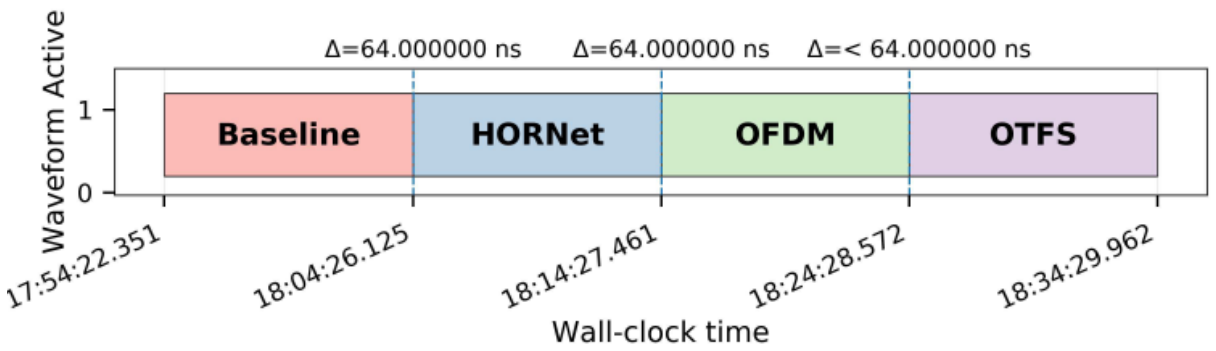

Fig. 10: StormWave seamless switching across waveforms.

| From | To | Previous End Time | Next Start Time | Gap ($\Delta$) |
|---|---|---|---|---|
| Baseline | HORNet | 18:04:26.124 | 18:04:26.125 | 64 ns |
| HORNet | OFDM | 18:14:27.461 | 18:14:27.461 | 64 ns |
| OFDM | OTFS | 18:24:28.572 | 18:24:28.572 | $<$ 64 ns |

TABLE I: Waveform transition timing derived from configuration timestamps. The gap $\Delta$ is computed as the difference between the start time of the next waveform and the end time of the previous.

**Waveform Switching.** We test the waveform switching capability of StormWave in this experiment. Fig. 10 illustrates the waveform switching strategy. As summarized in Table I, the measured transition gaps are 64 ns (Baseline→HORNet and HORNet→OFDM), and less than 64 ns (OFDM→OTFS). These sub-100 ns gaps are near the timestamp resolution and are thus negligible at the experimental time scale. This explains the visually continuous operation observed in the timeline and confirms StormWave's ability to support sub-microsecond seamless waveform switching without interrupting ongoing transmission, enabling rapid runtime reconfiguration during active operation.

**Dataset Collection.** We have collected two types of dataset during the experiments: metrics-based datasets and raw baseband I/Q datasets. Ground experiments log packet-level performance metrics, including throughput, access statistics, and configuration parameters, while A2A experiments record raw receiver I/Q snapshots due to mobility and synchronization constraints. The I/Q datasets enable post-processing using symbol- and distribution-level metrics such as ASER and KLD. The experiments span multiple sessions and link distances to comprehensively characterize the propagation conditions, with metadata recorded for each run to ensure traceability and reproducibility across experimental runs.

## IV. Conclusions

In this work, we have presented StormWave, an open, portable and extensible SDR-based RF interference platform for realistic evaluation of wireless communication systems. StormWave has been designed to enable intuitive configuration, seamless runtime waveform switching, and to support both built-in and user-defined interference waveforms. Ground-based and A2A experiments demonstrated controlled and repeatable interference under realistic propagation conditions, while seamless switching experiments confirmed sub-microsecond transition gaps during active transmissions. With these capabilities, StormWave is expected to provide a practical and extensible platform for studying interference resilience of wireless systems. In future work, we will extend StormWave toward adaptive and learning-driven generation of interference, and enable remote access through cloud-based testbed platforms such as UnionLab [20], [21].